\documentclass{article}

\usepackage{arxiv}

\usepackage[utf8]{inputenc} 
\usepackage[T1]{fontenc}    
\usepackage{hyperref}       
\usepackage{url}            
\usepackage{booktabs}       
\usepackage{amsfonts}       
\usepackage{nicefrac}       
\usepackage{microtype}      
\usepackage{lipsum}
\usepackage{graphicx}
\usepackage{float}

\title{Coevo: a collaborative design platform with artificial agents}

\author{
  Gerard Serra\\
   GTM-Grup de recerca en Tecnologies M\`{e}dia\\
  La Salle-Universitat Ramon Llull\\
  Barcelona, Spain \\
  \texttt{gerard.serra@salle.url.edu} \\
   \And
 David Miralles \\
  GTM-Grup de recerca en Tecnologies M\`{e}dia\\
  La Salle-Universitat Ramon Llull\\
  Barcelona, Spain \\
  \texttt{david.miralles@salle.url.edu} \\
}

\begin{document}
\maketitle

\begin{abstract}
We present Coevo, an online platform that allows both humans and artificial agents to design shapes that solve different tasks. Our goal is to explore common shared design tools that can be used by humans and artificial agents in a context of creation. This approach can provide a better knowledge transfer and interaction with artificial agents since a common language of design is defined. In this paper, we outline the main components of this platform and discuss the definition of a human-centered language to enhance human-AI collaboration in co-creation scenarios.
\end{abstract}


\section{Introduction}
Recent advances on Artificial Intelligence (AI) techniques have popularized their usage on a large number of applications. Specially on design and creative processes AI has emerged to support human design and creativity \cite{ha2017neural} \cite{roberts2018hierarchical} \cite{valenzuelarunway}. As stated on \cite{miikkulainen2019creative}, the next step in this field is machine creativity in order to augment human capabilities on problem solving. Specially, in the design process, \cite{maher2003co} proposed a computational and cognitive model assuming two parallel search spaces: the problem space and the solution space. Then an iterative process of search and evaluation within both spaces can allow to continually explore design possibilities. 

To address this topic, evolutionary computing \cite{eiben2003introduction}
has been demonstrated as a powerful tool to explore solution space and generate valid proposals \cite{lehman2018surprising}  \cite{lipson2000automatic} \cite{cully2015robots}. Many computational tools are been used to support human via generating or optimizing design solutions \cite{bentley1999evolutionary} \cite{kazi2017dreamsketch} \cite{philippaontology}. However, interaction with these systems is often based on providing some inputs and receiving output solutions generated by the system.
This differs with human design process on which there is an inherent learning. A design process influences on perception and knowledge acquisition due to the exploration of possible designs and the relationships with their context \cite{gero1996creativity}. Then, in a context of AI-human collaboration, this knowledge is lost if AI systems reasoning is not well-communicated and understood by humans.

Following this idea, other research lines seek a more interactive and intelligible collaboration in this context of creation \cite{fails2003interactive} \cite{oh2018lead}. We highlight the work presented by \cite{guzdial2019friend} where a human designer collaborates with multiple artificial agents to co-design a game level.
We consider that in order to support humans in a creative context, it is crucial to explore techniques that enable better knowledge transfer between humans and artificial agents during the whole process of designing rather than only receiving final outputs given by AI systems.

In this paper, as a first step, in order to explore this AI-human collaboration, we propose an environment where both artificial agents and humans can learn to design solutions for an specific problem.

\section{Concept design}

In Coevo, we have defined a shared environment for both artificial agents and humans. There, we propose a common set of creation tools establishing the same language of design and common evaluation criteria on design proposals.  This approach can benefit knowledge transfer since all the actions taken during the process of creation can be interpreted. 

\begin{figure}[H]
  \centering
  \includegraphics[width=1\textwidth]{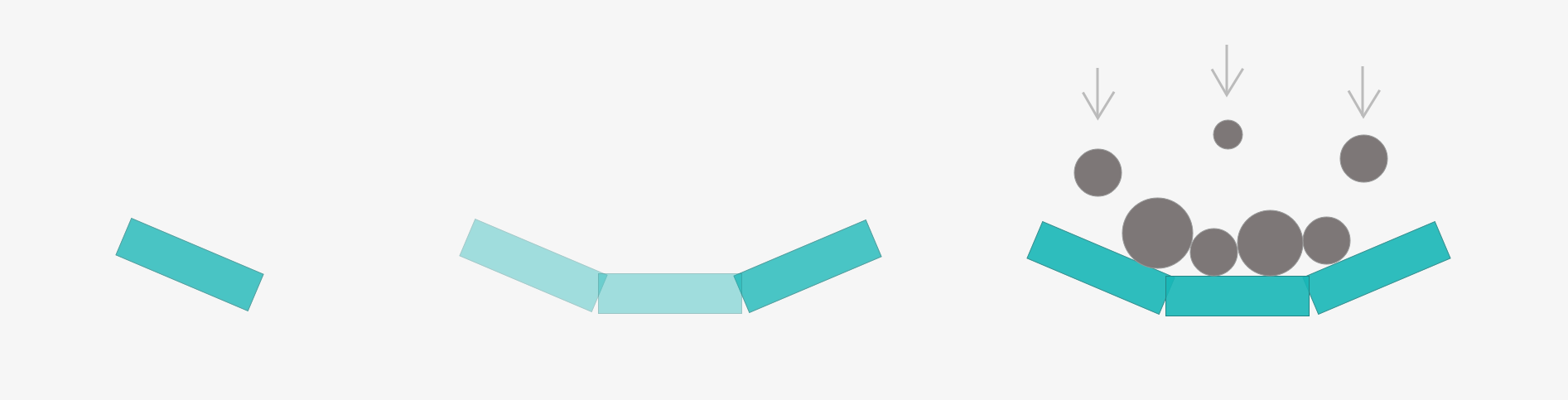}
  \caption{2D shape emerges by combining multiple 2D pieces. This example illustrates the creation of a bowl in our environment}
  \label{fig:Shape}
\end{figure}
\subsection{Interactive tools}

As shown on Figure \ref{fig:Shape}, design proposals are composed by a set of 2D bricks assembled together as a continuous chain that generates a 2D shape. Our interaction model is based on manipulating this common set of simple building blocks that, when combined together, a more complex structure can emerge. These structures are created by addition and subtraction of blocks and via rotating them in any direction defining the direction. This set of actions to be performed is the same for both artificial agents and humans. Moreover, all the actions taken during the process of creation are saved. This enables to access and visualize not only the final proposals but also all the design process until arriving to a certain final decision.

\subsection{Environments}

\begin{figure}[h]
  \centering
  \includegraphics[width=\textwidth]{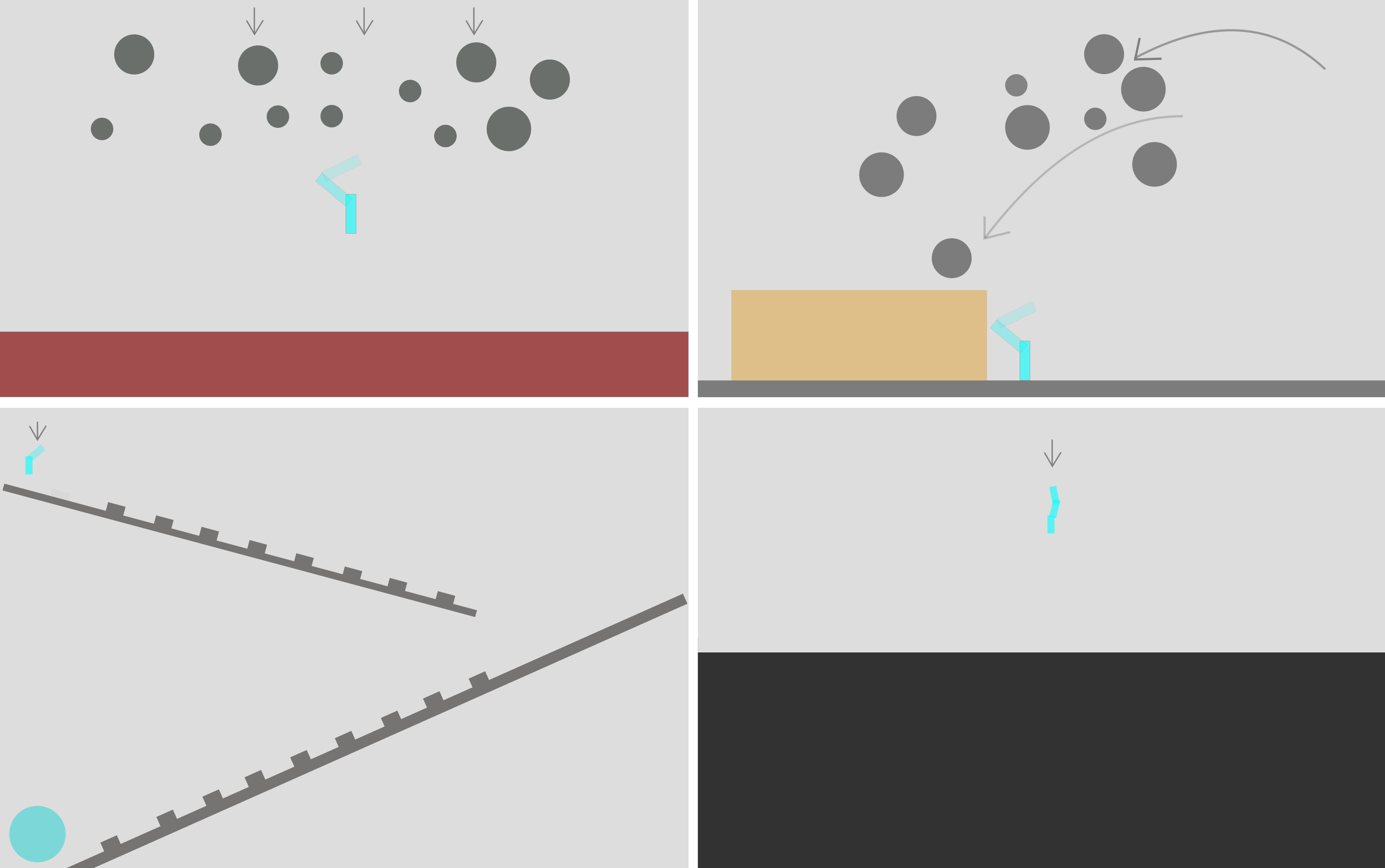}
  \caption{Design challenges. From top-left to bottom-right: Collect falling balls; Protect orange area; Move to certain point;  Cut through a dense medium (dark area) }
  \label{fig:experiments}
\end{figure}

In order to allow wider audiences to interact with our proposal we decided to create an online platform. There, users can generate their design proposals in our collection of environments. These environments are based on Javascript \cite{flanagan2006javascript} libraries p5.js and matter.js \cite{mccarthy2017p5} combined together to generate an interactive web interface with capabilities to run 2D rigid body physics. As show on Figure \ref{fig:experiments}, a total of four design challenges have been defined: collect, move, cut and protect. For each challenge, an environment with initial design conditions is generated, involving positioning of the elements and design proposal and their physical behavior.  In addition, an objective definition is set in order to measure design performance.

When human or artificial agent propose a design, the system will place it on the environment and evaluate it regarding each objective definition. Then it returns an score from 0 to 1 to quantify the overall performance within that experiment based on each specific goal definition. 

Similar to other research projects and benchmarks \cite{bellemare2013arcade} \cite{mnih2015human} \cite{brockman2016openai}, by evaluating within the same conditions an agent behaviour we can directly compare their performance with humans one.  Since our platform provides a quantitative score based on design performance, an incremental learning process can be defined. Designers will be encouraged to obtain output solutions with their higher score associated. Same goes for our algorithm that by receiving the scores of the generated proposals it will seek to maximize its final score.

Apart from that, given a certain challenge there is the possibility that many different solutions are valid.  A richer solution space will emerge when several valid proposals appear, so we want to maximize it.  

Moreover, since the creation tools are common, at any point of the process we can visualize the experiments and understand how a proposal is constructed. This can allow editing in any step and enable capabilities to retake the experiments.

\subsection{Artificial agent definition}
We want to explore how multiple proposals can emerge given a specific design challenge. For that reason, our algorithm starts designing from scratch rather than using labeled data with possible solutions to train our system. Following this approach, we propose to use a general common algorithm for all the experiments. We decided to create a  population-based optimization algorithm that learns using the scores obtained when system evaluates each design proposal. 

In contrast to many design optimization techniques, we propose an interactive process allowing users to visualize all the steps in the optimization. Moreover, since the design tools are the same, we potentially allow users to contribute at any point of the optimization, fine-tuning any piece that composes the 2D shape and continue the learning process without relaunching the experiment. This also helps to increase system transparency which becomes more explainable to humans during the whole process (the opposite of a black box algorithm). We can measure intelligibility of an artificial system with possible interactions and potential contributions that an user can do in any step of its learning process. 

\begin{figure}[h]
  \centering
  \includegraphics[width=1\textwidth]{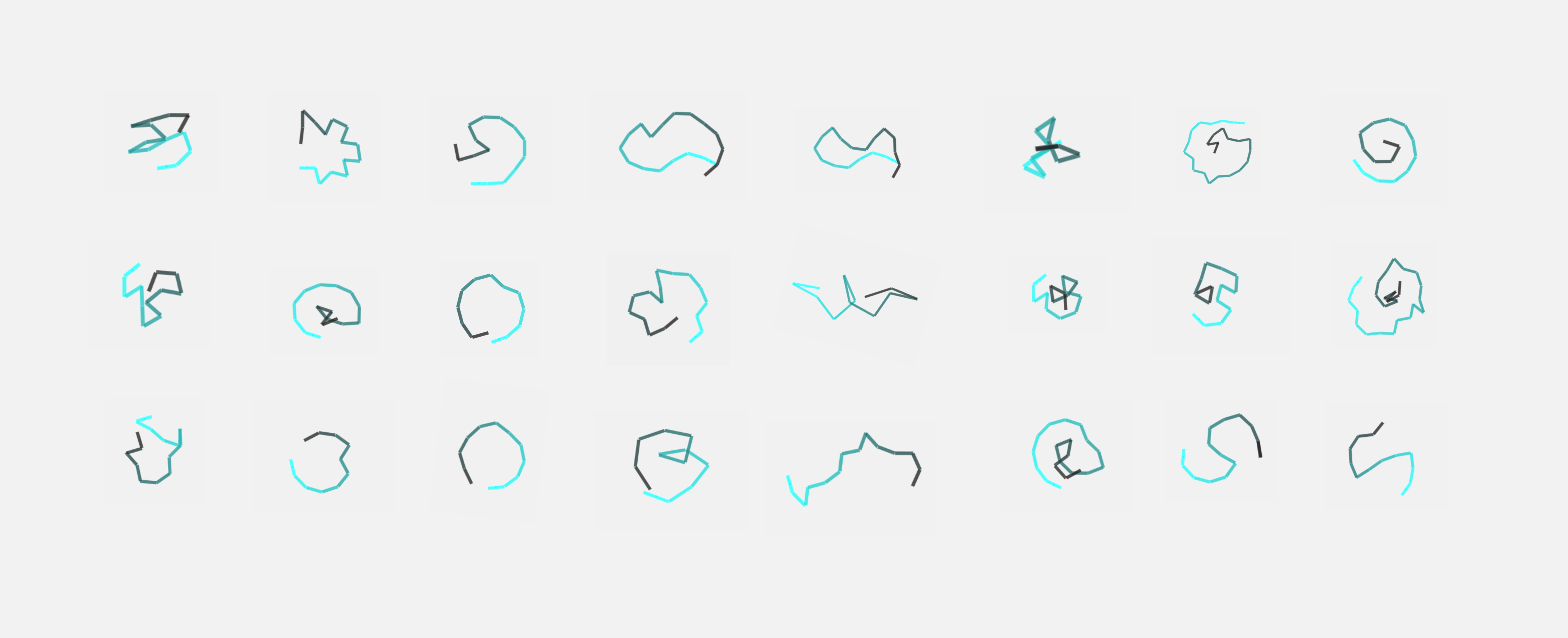}
  \caption{Gallery of AI generated designs. In a design problem such as "Create an object that moves on a inclined plane" we can observe multiple solutions that differ of the common solution in that context: the wheel}
  \label{fig:wheels}
\end{figure}

Finally, as described in \cite{miikkulainen2019creative}, population-based algorithms can explore wider solution spaces and potentially propose novel designs. These unexpected proposals can inspire creators with new approaches to solve a defined problem and augment their capabilities on designing by expanding their vision on a possible solution space \cite{lehman2018surprising}. As an example, in Figure \ref{fig:wheels} we expose valid solutions generated by our algorithm when given the task to create an object that moves through an inclined plane. 

\section{Discussion and future work}
In this paper, we discussed how a common environment and tools can support user understanding of artificial agents decisions and enable collaborations within a full process of design. 

In particular, we presented a proof-of-concept scenario  where both artificial agents and humans create 2D shapes to perform a certain task in multiple physically simulated environment. Building complex structures with minimal elements (2D pieces) and the usage of simple actions (rotate/add/remove) allows a wide range of creation possibilities. 
Moreover, common creation tools allow better knowledge transfer, compare proposals and explore on new designs. Population-based optimization algorithms also help on maximizing the exploration of the solution space. This exploration provides not only variants of similar designs but also sometimes radically novel proposals. This new knowledge can help to augment human design capabilities and offers a broader view of the problem and the solution space.

Finally, we consider that our proposal opens a new space to explore future scenarios of co-designing with artificial agents encouraging wider audiences to interact and design with AI.

\section{Acknowledgements}
This work has been partially funded with the support of the European Social Fund and the Secretaria d'Universitats i Recerca del Departament d'Economia i Coneixement of the Catalan Government for the pre-doctoral the pre-doctoral FI grant No. 2018FI\textunderscore 01052.

\bibliographystyle{unsrt}

\end{document}